\documentclass{optica-article}

\journal{opticajournal} 

\articletype{Research Article}

\begin{document}

\title{Propagation of pulsed light in an optical cavity in a gravitational field}

\author{Daniel D. Hickstein\authormark{1}, David R. Carlson\authormark{1}, Zachary L. Newman\authormark{1}, Cecile Carlson\authormark{1}, Carver Mead\authormark{2}\authormark{*}}

\address{\authormark{1}Octave Photonics, 325 W South Boulder Road, Louisville, CO 80027 USA\\
\authormark{2}California Institute of Technology, Pasadena, CA 91125, USA\\}

\email{\authormark{*}carver@caltech.edu} 

\begin{abstract*} 
All modern theories of gravitation, starting with Newton's, predict that gravity will affect the speed of light propagation. Einstein's theory of General Relativity famously predicted that the effect is twice the Newtonian value, a prediction that was verified during the 1919 solar eclipse. Recent theories of vector gravity can be interpreted to imply that gravity will have a different effect on pulsed light versus continuous-wave (CW) light propagating between the two mirrors of an optical cavity. Interestingly, we are not aware of any previous experiments to determine the relative effect of gravity on the propagation of pulsed versus CW light. In order to observe if there are small differences, we use a 6 GHz electro-optic frequency comb and low-noise CW laser to make careful measurements of the resonance frequencies of a high-finesse optical cavity. Once correcting for the effects of mirror dispersion, we determine that the cavity resonance frequencies for pulsed and CW light are the same to within our experimental error, which is on the order of $10^{-12}$ of the optical frequency, and one part in 700 of the expected gravitational shift. 
\end{abstract*}

\section{Introduction}
Einstein first predicted\cite{einstein1907} that the frequency of a clock ($\omega$) would be proportional to the Newtonian gravitational scalar potential ($\Phi$) as $\omega\propto(1+\Phi/c_0^2)$, where $c_0$ is the speed of light far from any mass concentration.  He followed this conclusion directly with an argument that the speed of light $c_1$ at gravitational potential $\Phi$ would be $c_1=c_0(1+\Phi/c_0^2)$. His 1912 theory\cite{einstein1912} concluded that the gravitational potential would have both a scalar and vector character, but he never applied that insight to the propagation of light. By 1915, his General Relativity (GR) theory\cite{einstein1915} concluded that the gravitational potential would have \textit{twice} the effect on the speed of light as on the rate of clocks; the speed of light would become $c_2=c_0(1+2\Phi/c_0^2)$. Both the rate of atomic clocks \cite{chou2010, herrmann2018, bothwell2022} and the speed of light \cite{shapiro1964, shapiro1968, Demorest_2010, dyson1920} have been experimentally observed as functions of gravitational potential, and the results agree with the predictions of GR. The physical origin of the factor of two between the two effects is described by Einstein \cite{Einstein1922, Thorne_1994}: Since atomic frequencies are proportional to $c_1$, and the speed of light is proportional to $c_2$, and it is also \textit{postulated} that the frequencies of atomic clocks and optical cavity resonances stay equal as the gravitational potential changes, and the frequency of an optical cavity of length $l$ is $f_{\rm opt}=c_2/l$, then $l$ must be proportional to $c_1$.  No physical basis for this dependence is given---it still remains a mystery.

Recently one of us (CM) has expanded upon Einstein's 1912 suggestion \cite{einstein1912} that gravitational potential has both scalar and vector components, and introduced G4v\cite{mead2023, mead2023b}, a theory of gravitation based the quantum-wave nature of matter, and squarely incorporating Einstein's insight. G4v gives the same results as GR for the historical experimental tests, including the precession of the perihelion of Mercury\cite[Ch.~10]{mead2023}, and the geodetic effect of Gravity Probe B \cite[Ch.~11]{mead2023}. Moreover, a simple G4v cosmology\cite{mead2023b}, fit to recent supernovae observations\cite{Riess_2011}, gives Mach's Principle, while avoiding the requirement for dark energy. Straightforward deductions within the G4v framework lead to the conclusion that both the scalar and vector terms each introduce equal contributions to the slowing of light propagation near massive objects \cite[Ch.~7]{mead2023}. In this theory, the sum of the two effects accounts for the factor of two in an intuitive manner. 

One method to experimentally test this theory is to compare a situation where light is affected by both the scalar and vector potentials with a situation where light is only affected by the scalar potential. Light pulses propagating in one direction are an obvious example of a situation where the vector and scalar potentials will both contribute and the resonances would follow $c_2$. Finding a situation where light is affected only by the scalar potential is more difficult. From one point of view, a standing-wave, resonant in an optical cavity, has no net momentum, and would thus be exempt from the vector part of the slowing, and the cavity continuous-wave (CW) resonances would follow $c_1$. The contrary argument holds that a standing-wave is simply the sum of two independent counter-propagating traveling-waves, each launched from its respective mirror, and each would be fully affected by the vector coupling to matter. From this second point of view, there should be no difference in the gravitational slowing of light between traveling-waves and standing-waves in an optical cavity. Since one can arrive at either outcome by a principled argument, an experiment was obviously necessary to settle the question, and to guide the course of future investigation.

\begin{figure}[ht!]
\centering\includegraphics[width=4in]{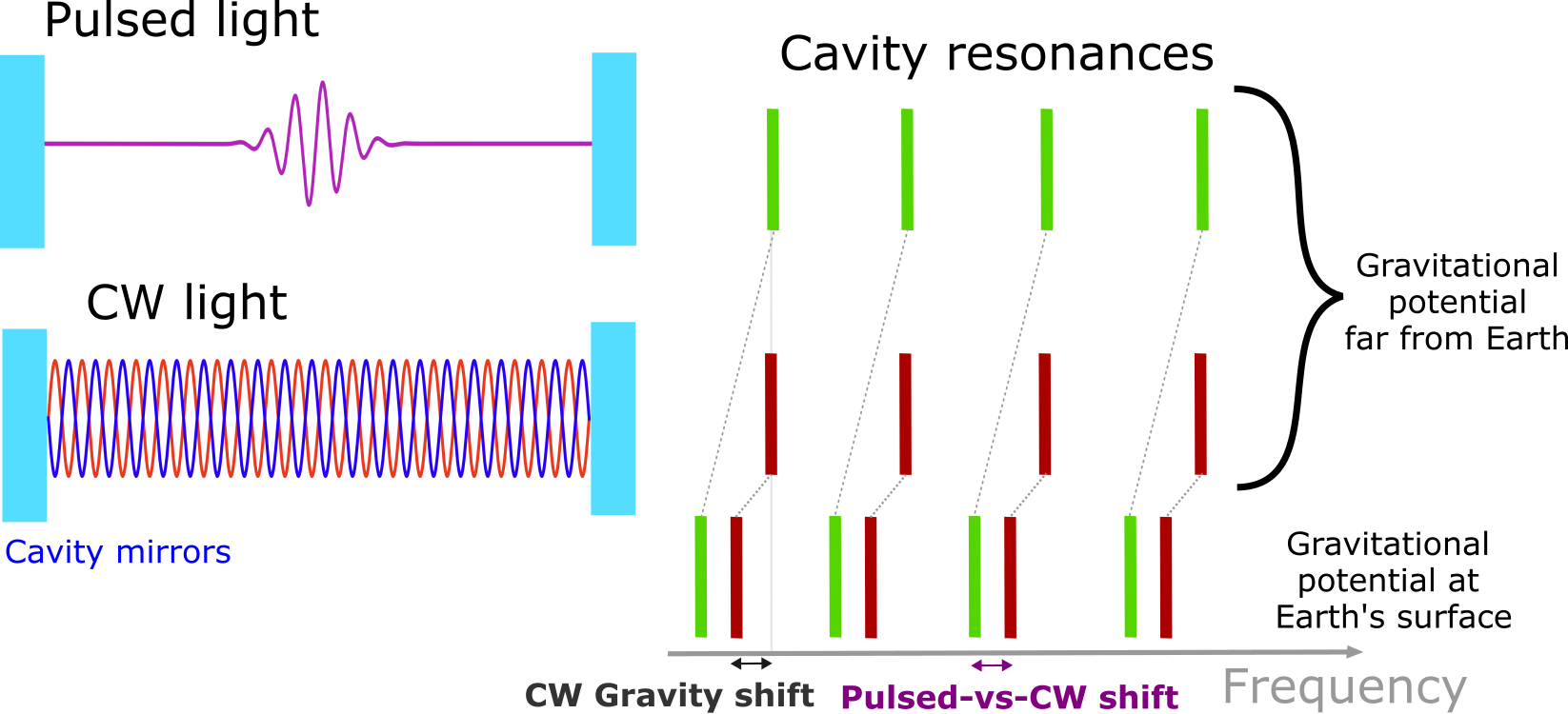}
\caption{\label{fig:concept} (left) Both pulsed and continuous wave (CW) light can propagate between two highly reflective mirrors in an optical cavity. (right) When the optical cavity is brought into a different gravitational potential, the cavity resonance frequencies will shift. It is possible that gravity has a different effect on pulsed light versus CW light, resulting in a shift between the location of the cavity resonances when the cavity exists in a gravitational potential well, such as on the surface of the Earth. Our experiment attempts to detect if such a ``Pulsed-vs-CW shift'' exists.}
\end{figure}

Thus, we have conducted the first (to our knowledge) experiment to determine if an optical cavity has precisely the same resonant frequencies for pulsed versus CW light. In our experiment, both a standing-wave and a pulse of light circulate between the same two mirrors of the same optical cavity (Figure~\ref{fig:concept}). The gravitational scalar and vector potentials on Earth's surface are both lower than those far from Earth by a factor of $MG/{c_0}^2R \approx 6.9\times 10^{-10}$, where $M$ is the mass of the Earth, $G$ is the gravitational constant, and $R$ is the radius of the Earth \cite{mead2023}. If gravity affects the propagation of pulsed and CW light differently, their respective resonances should be shifted with respect to each other, as shown in the lower section of Figure~\ref{fig:concept}. The use of highly reflective mirrors in the optical cavity provides narrow cavity resonances and allows us to achieve a very high level of precision.

\section{Experiment}
\begin{figure}[ht!]
\centering\includegraphics{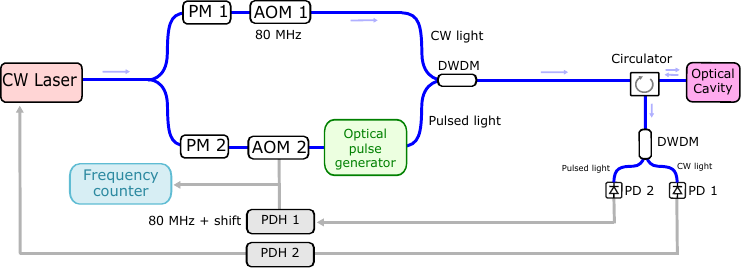}
\caption{\label{fig:experiment} The experimental concept for detecting the difference in cavity resonance frequencies between pulsed and CW light. A CW laser is split into two arms, with the top arm remaining CW while the bottom arm is converted into picosecond pulses via a optical pulse generator. Two phase modulators (PM~1 and PM~2) are used to apply modulation for Pound-Drever-Hall (PDH) locks to the optical cavity. One PDH lock applies feedback to the frequency of the CW laser to lock it to the optical cavity. The first acousto-optic modulator (AOM~1) applies a constant 80~MHz frequency shift, while AOM~2 applies a variable frequency of approximately 80~MHz to lock the pulsed light to the optical cavity. This ${\sim}80$~MHz signal is then passed to the frequency counter, which measures the exact frequency. This frequency represents the experimental pulsed-versus-CW shift plus 80~MHz. Dense wavelength division multiplexers (DWDMs) are used to split and recombine the pulsed and CW light, and InGaAs photodetectors (PDs) are used for detection.}
\end{figure}

In order to test if an optical cavity presents the same resonance frequencies when probed with CW and pulsed light, we begin with a single CW laser and split the light into two arms (Figure~\ref{fig:experiment}), where one arm will remain CW and the other will be converted into picosecond pulses using an ``optical pulse generator''. The optical pulse generator starts with an electro-optic frequency comb (EO comb)\cite{kobayashi1972, kourogi1993, carlson2018}, which uses electro-optic phase and intensity modulators to carve the CW light into ${\sim}50$ ps chirped pulses. Then, a grating-based compressor\cite{treacy1969} provides dispersion compensation to compress the pulses to ${\sim}2$ ps. Finally, an optical amplifier increases the pulse energy (Figure~\ref{fig:experiment_detailed}). The pulse durations are characterized using a frequency-resolved optical gating (FROG)\cite{trebino1997} device (MesaPhotonics VideoFROG). A fiber-based dense wavelength division multiplexer (DWDM) is used to suppress the spectral region near the center of the pulsed light, thereby providing spectral separation between the pulsed and CW light. Since it clips the spectrum, the DWDM causes the pulses to become somewhat longer in time, and exhibit additional temporal structure. However, the resulting ${\sim}5$ ps pulses are still short compared to the cavity round-trip time of 167~ps (Supplementary Figures~\ref{fig:frog_full} and~\ref{fig:frog_dwdm}).

The two arms are then combined and delivered to an ultrastable optical cavity. The optical cavity (Stable Laser Systems) consists of two highly reflective mirrors (cavity finesse of ${\sim}$ 250,000 at 1550 nm) separated by a monolithic cubic spacer made from low-thermal-expansion glass. The cavity is actively temperature controlled and housed within a vacuum chamber. Since the cavity is in vacuum, the main contributions to the cavity mode frequencies will be due to the separation between the mirrors and the dispersion of the mirror coatings \cite{siegman1986}.

Each arm contains a phase modulator (PM 1 and PM 2), which provides the phase modulation required for the Pound-Drever-Hall (PDH) method \cite{drever1983} of laser-cavity stabilization. A second DWDM is used to spectrally separate the CW and pulsed light and direct them onto separate photodetectors. The PDH signal for the CW light is connected to a servo controller (Vescent D2-125), which provides feedback to the CW laser frequency. The PDH signal for the pulsed light is used to servo the frequency of an acousto-optic modulator (AOM) that shifts the frequency of the CW light prior to the optical pulse generator. A separate AOM provides a fixed frequency shift (80 MHz) in the CW arm. 

In this manner, the frequency used to drive the variable-frequency AOM (AOM~2 in Fig.~\ref{fig:experiment}) represents the shift between the cavity mode interrogated by the CW laser and the nearest comb tooth of the pulsed laser (plus 80~MHz). Thus, the frequency used to drive AOM~2 (minus 80~MHz) is a direct probe of the difference in the cavity response to the broadband pulsed light and the CW light. This observed shift will be a sum of the effect of cavity-mirror dispersion and effects resulting from the pulsed versus CW nature of the light. The error in the measurement of the observed shift is largely determined by systematic uncertainties, primarily the precise voltage of the PDH lockpoint. By making small adjustments to the PDH lockpoint, we estimate this error to be approximately 200~Hz (standard deviation). Further experimental details are provided in Section~\ref{ssec:experimental_details} of the Supplementary Materials.

In order to interpret the observed shifts, we must account for the dispersion of the optical cavity mirrors. The group-delay dispersion (GDD) of the mirror coatings causes the frequencies of the cavity modes to deviate from a perfect grid \cite{thorpe2005}. For the case of constant GDD, the cavity modes will become sequentially more closely spaced on one side of the center frequency and more widely spaced on the other side. The result is that broadband light will experience a shift in the $average$ mode location that depends quadratically on the bandwidth of the pulse\cite{paschotta2023}. Thus, in the absence of other shifts, we expect to see a quadratic dependence on the shift with increasing bandwidth for pulsed light for a given laser center frequency. By measuring the shift as a function of bandwidth, we can estimate the contribution of the mirror GDD. By repeating this experiment with lasers of different center wavelength, we can explore different values of mirror GDD, helping to separate the effects of mirror GDD from those of gravity.   

Of course, we would also expect any gravity-related pulsed-vs-CW shift to also decrease as we decrease the bandwidth of our optical pulses. In other words, as we make the pulses progressively longer, we would expect them to eventually converge to the CW case. However, in our experiments we vary the bandwidth over the range of 60 to 119 GHz, which is significantly larger than the cavity mode spacing of 6 GHz. Therefore, as long as we achieve reasonable compression of the pulses, their durations will always be much shorter than the cavity round-trip time. Thus, we can, to a first approximation, assume that any pulsed-vs-CW shift resulting from the gravitational potential is constant for the bandwidth explored in this study.

To explore the role of cavity mirror dispersion we utilize lasers at three different wavelengths: 1558.6, 1552.5, and 1529.6 (hereafter referred to as  1559, 1553, and 1530 nm) in order to explore different dispersion regimes of the cavity mirrors. We note that, due to insufficient amplifier gain at shorter wavelengths, we were not able to perform a pulse duration measurement of the 1530 nm pulses. Nevertheless, we expect these pulses are still short compared to the cavity round-trip time.

\vfil\eject
\section{Results and Discussion}
\begin{figure}[ht!]
\centering\includegraphics[width=5.5in]{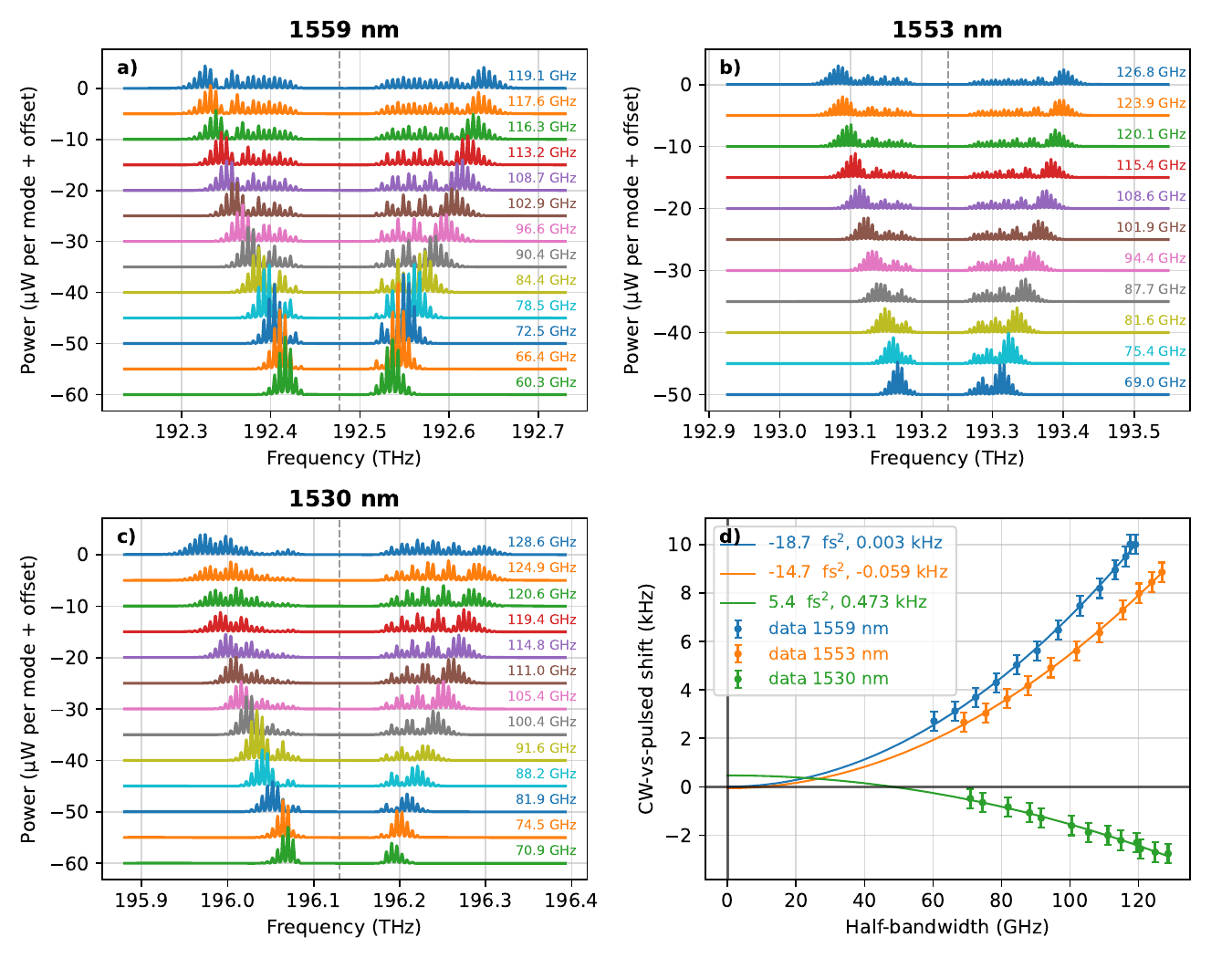}
\caption{\label{fig:shift_vs_bandwidth} The experimentally observed shift between CW and pulsed light is strongly dependent on the bandwidth. a,b,c) The optical spectra of the pulsed light for the  $1559\ \mathrm{nm}$, 1553 nm, and 1530 nm driving lasers. d) The shift between pulsed and CW light as a function of the bandwidth. A quadratic fit can be used to extract the effective group-delay dispersion in $\mathrm{fs}^2$.}
\end{figure}

We measured the shift between pulsed and CW cavity resonance positions as a function of the comb's optical bandwidth by decreasing the 6 GHz drive power to the phase modulators. We repeated this experiment repeated for the three laser wavelengths (Figure~\ref{fig:shift_vs_bandwidth}a-c). The observed shift as a function of bandwidth follows a clear quadratic trend (Figure~\ref{fig:shift_vs_bandwidth}d). By fitting the data using a quadratic function plus a vertical offset, we can extract the GDD of the cavity mirrors as well as the residual pulsed-vs-CW shift that is not explained by the mirror GDD. 

\begin{figure}[ht!]
\centering\includegraphics[width=3in]{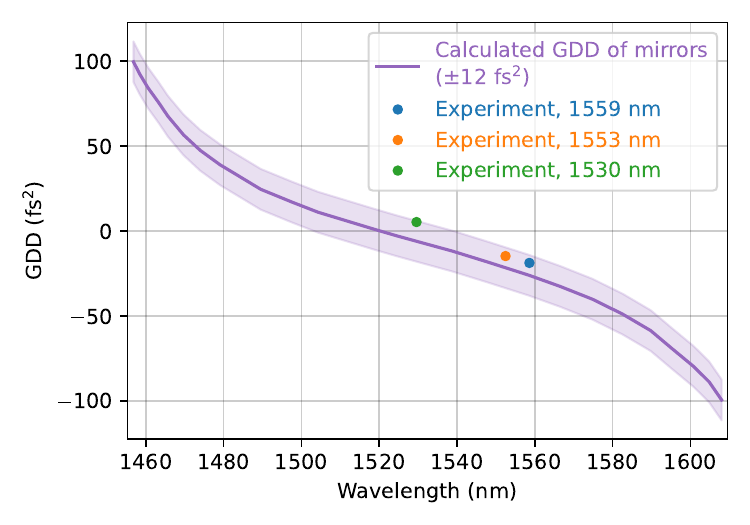}
\caption{\label{fig:cavity_dispersion} The calculated GDD of the mirrors compared with the values extracted from the shift-versus-bandwidth measurements (Figure \ref{fig:shift_vs_bandwidth}). We expect the true dispersion of the mirrors (purple region) to be within a few $\mathrm{fs}^2$ of the calculated design dispersion (purple curve). The values extracted from our measurements (points) are slightly above the calculated dispersion curve, but within the expected error.}
\end{figure}

We find that two of the lasers (1559~nm and 1553~nm) are in the anomalous dispersion regime, while the 1530 nm laser is in the normal dispersion regime (Figure~\ref{fig:shift_vs_bandwidth}d). We compare our extracted dispersion values to the calculated curve for the mirror design and find reasonable agreement (Figure~\ref{fig:cavity_dispersion}). Our values are consistently higher than the theoretical curve by approximately 7~$\mathrm{fs}^2$, implying that the true mirror dispersion is somewhat shifted from the calculated curve. Such a deviation could result from small deviations in the thickness of the mirror coatings from their design values.


Having confirmed that the cavity mirror GDD affects the experimental pulsed-vs-CW shift in a predictable way, we can then determine the magnitude of the shift that is not explained by the GDD. This residual pulsed-vs-CW shift is seen as the vertical offset between the fit lines and the origin of the plot in Figure~\ref{fig:shift_vs_bandwidth}d. This offset is found to be 3 Hz, -59 Hz, and 473 Hz for the 1559 nm, 1553 nm, and 1530 nm lasers respectively. As discussed previously, systematic uncertainties limit our measurement of the shifts on the level of approximately 200~Hz. Thus, our residual pulsed-vs-CW shifts are either within the error to zero or nearly so.

\begin{figure}[ht!]
\centering\includegraphics[width=3.5in]{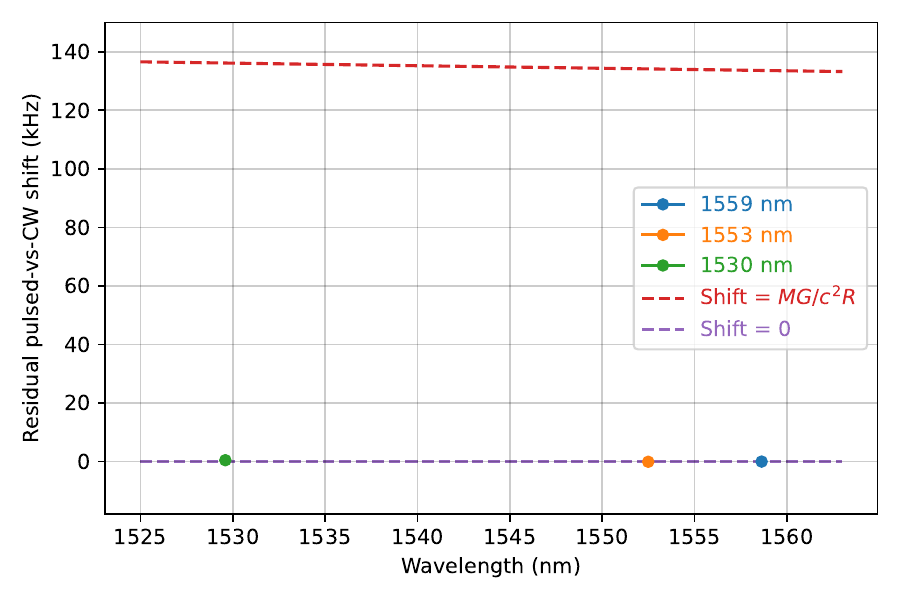}
\caption{\label{fig:gravity_shift} Residual pulsed-vs-CW shift with the effect of cavity-mirror dispersion removed. The dashed lines represent the expected results for the situation where the effect of gravity causes a shift proportional to $MG/(c^2R)$ (top line) and the situation where the cavity resonances are the same for pulsed and CW light (Shift=0). The error bars are smaller than the data points, supporting the identical propagation of pulsed and CW light.}
\end{figure}

The residual pulsed-vs-CW shift for each wavelength is shown in Figure~\ref{fig:gravity_shift}, which helps to visualize the difference between the observed residual pulsed-vs-CW shifts and the predicted shifts. These results indicate that CW and pulsed light interact with an optical cavity in the same way to within an uncertainty of approximately 200~Hz, a fractional uncertainty $10^{-12}$ of the optical frequency of ${\sim}200$~THz. Since the anticipated gravitational pulsed-versus-CW shift is $6.9\times10^{-10}$ (133 kHz for a 1553 nm laser), we can rule out this prediction by a factor of 700.

\vbox to 2in{}
\vfil\eject

\section{Conclusion}
Here we explored the proposal that a gravitational potential may affect pulsed and CW light propagation velocity differently, due to the vector coupling of the light with the gravitational potential. To test this hypothesis, we constructed an apparatus consisting of an ultra-stable optical cavity simultaneously probed with both CW and pulsed lasers. While small differences in the cavity resonance frequencies were observed, these differences can be attributed to the group-delay dispersion of the mirror coatings. Thus, we conclude that optical cavities respond identically to pulsed and CW light to within our experimental error, which is approximately one part in 700 of the gravitational shift, and one part in $10^{12}$ of the optical frequency. This result suggests that pulsed and standing-wave light experience identical velocity shifts in a gravitational potential. 

We conclude that the proposition that pulsed and CW light experience a different shift in a gravitational potential\cite{mead2023} is incorrect. Thus, this experiment cannot provide the insight into the origin of the factor of two between speed of light and atomic reference dependencies on gravitational potential. A new level of understanding must be obtained to reflect this finding, and theories of vector gravity (such as G4v) must be formulated (or modified) to incorporate this understanding. 

Moreover, the conclusions of this experiment suggest another experiment. General Relativity \cite{Einstein1922} postulates that the length of an optical cavity is proportional to $c_1$, partially canceling the effect of the speed-of-light change and making cavity resonance and atomic frequency both have the dependence $c_1$. No physical basis for this GR assumption is presently known, and (to the authors' knowledge), there has been no experimental test of this hypothesis. However, this claim can be experimentally tested by comparing the resonance of an optical cavity with that of an atomic clock in a varying gravitational potential. Of course, because the reasonably achievable changes in gravitation potential resulting from moving to a higher elevation on Earth are extremely small, such an experiment would require a state-of-the-art portable optical atomic clock as well as an exceptionally stable optical cavity. However, recent advances in both portable optical atomic clocks \cite{roslund2024} as well as crystalline optical cavities \cite{kedar2023, milner2019, zhang2017} will allow for cavity--clock comparisons with unprecedented precision, enabling new tests of theories of gravitation.

\vfil\eject
\section{Supplemental Materials}

\subsection{Experimental details}
\label{ssec:experimental_details}
\begin{figure}[ht!]
\centering\includegraphics{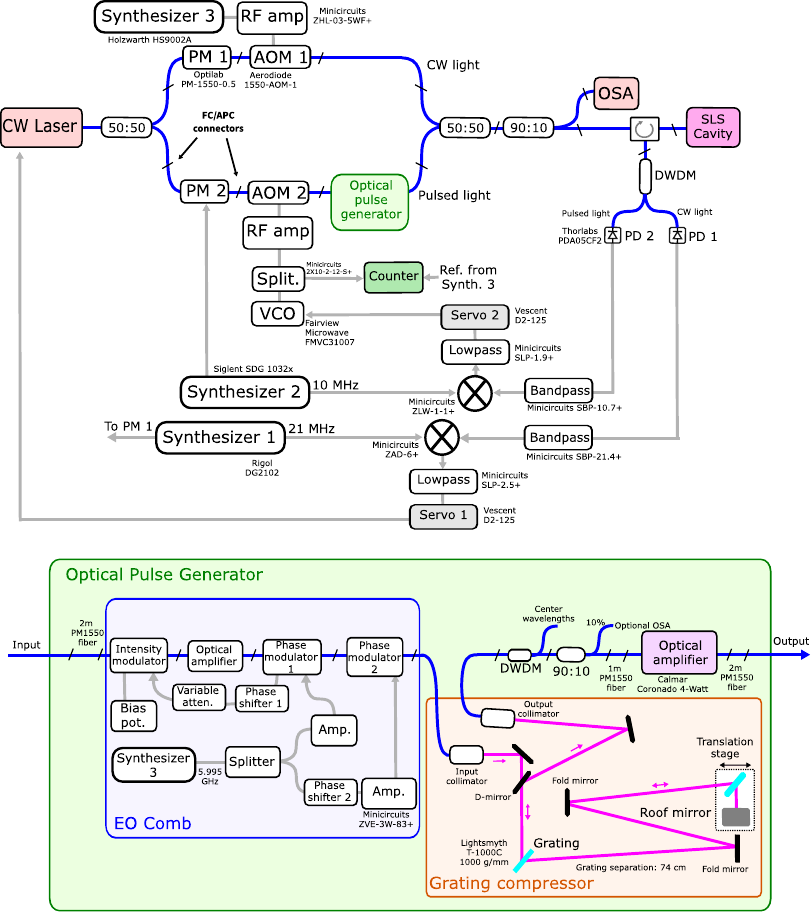}
\caption{\label{fig:experiment_detailed}A detailed view of the experimental apparatus. A CW laser is split into two arms, one remains CW and the other is converted into pulses via the optical pulse generation. The two arms are recombined before being locked to the high-finesse cavity via Pound-Drever-Hall locks.}
\end{figure}

The experimental concept (Figure~\ref{fig:experiment_detailed}) is to lock a 6 GHz pulsed laser to the resonances of a high-finesse cavity and compare the frequency of the central comb tooth to that of a CW laser locked to the same cavity. This will determine if the cavity resonances are the same for CW and pulsed light, or if they are different.

\vfil\eject

The experiment utilized three different stable CW lasers. The laser at 1558.6~nm (referred to as 1559~nm) is an OEWaves OE4040 Hi-Q laser and utilizes an built-in micro-resonator for enhanced stability. The laser at 1552.5 nm (referred to as 1553~nm) is a ULN15TK ultra-stable laser from Thorlabs. The laser at 1529.6~nm (referred to as 1530~nm) is an NP Photonics Rock erbium-fiber laser. 

The CW laser light is split into two paths, with the upper path remaining CW while the bottom path is converted into pulsed light via the optical pulse generator. The two beams are recombined and are then incident on the high finesse cavity. The reflected light from the cavity is split in a dense wavelength division multiplexer (DWDM, FiberMart) and the CW light and pulsed light are directed to separate 150~MHz InGaAs photodiodes (Thorlabs PDA05CF2). The RF signals from the photodiodes are demodulated in the standard Pound-Drever-Hall (PDH) \cite{drever1983} configuration. 

The PDH error signal from the CW laser is fed into a fast electronic servo box (Vescent D2-125) which provides feedback to lock the CW laser frequency to the optical cavity. The PDH error signal from the comb is fed into a second servo which provides feedback to the voltage-controlled oscillator (VCO) which is driving AOM2 at approximately 80~MHz. This AOM provides a tunable frequency shift to the pulsed laser to lock the comb frequencies to the optical cavity. AOM1 provides a fixed 80~MHz frequency shift. 

The optical pulse generation module consists of an electro-optic frequency comb (Octave Photonics GHz Electro-Optic Comb, GECO), a grating-based pulse compressor for dispersion compensation, and an optical amplifier (Calmar Coronado). The EO-Comb utilizes one intensity modulator and two phase modulators to produce chirped pulses. The grating compressor utilizes two 1000~groove/mm gratings with a separation of approximately 74~cm to apply the appropriate dispersion to compress the pulses. The light is then coupled back into fiber and passes through tow DWDMs, which filter out a region near the center of the spectrum and allows for the subsequent combination and separation of the pulsed and CW light. Two DWDMs are used to ensure 30+~dB suppression of the pulsed light in the center region.

\subsection{Spectra}

\begin{figure}[ht!]
\centering\includegraphics[width=5in]{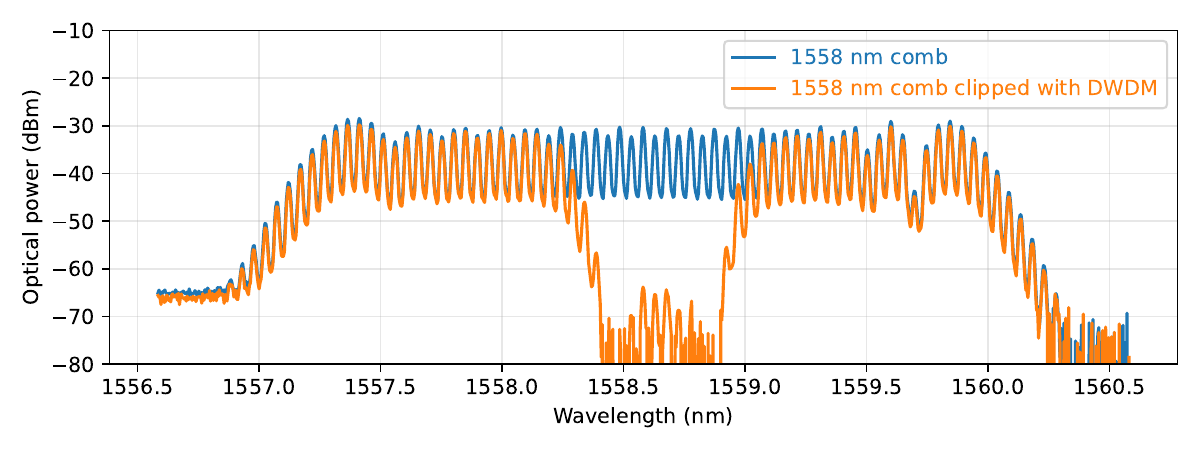}
\caption{\label{fig:specta1558} The spectrum of the 1558~nm comb before (blue) and after (orange) the center of the spectrum is removed using two DWDMs. The use of two DWDMs assures strong supression of the center comb modes.}
\end{figure}

The pulses were generated with approximately 3 nm of bandwidth, or roughly 70 comb teeth with 6~GHz spacing (Figure \ref{fig:specta1558}). The spectrum was then clipped using two DWDMs to remove the center portion. While this caused the pulse durations to increase, it was necessary in order to allow the pulsed and CW light to be separated onto different detectors for the PDH locking. Two DWDMs were used in order to achieve >30 dB suppression of the comb teeth in the center region to prevent any pulsed light from interfering with the PDH signal on the CW detector.

\subsection{Pulse duration}
\begin{figure}[ht!]
\centering\includegraphics[width=2.5in]{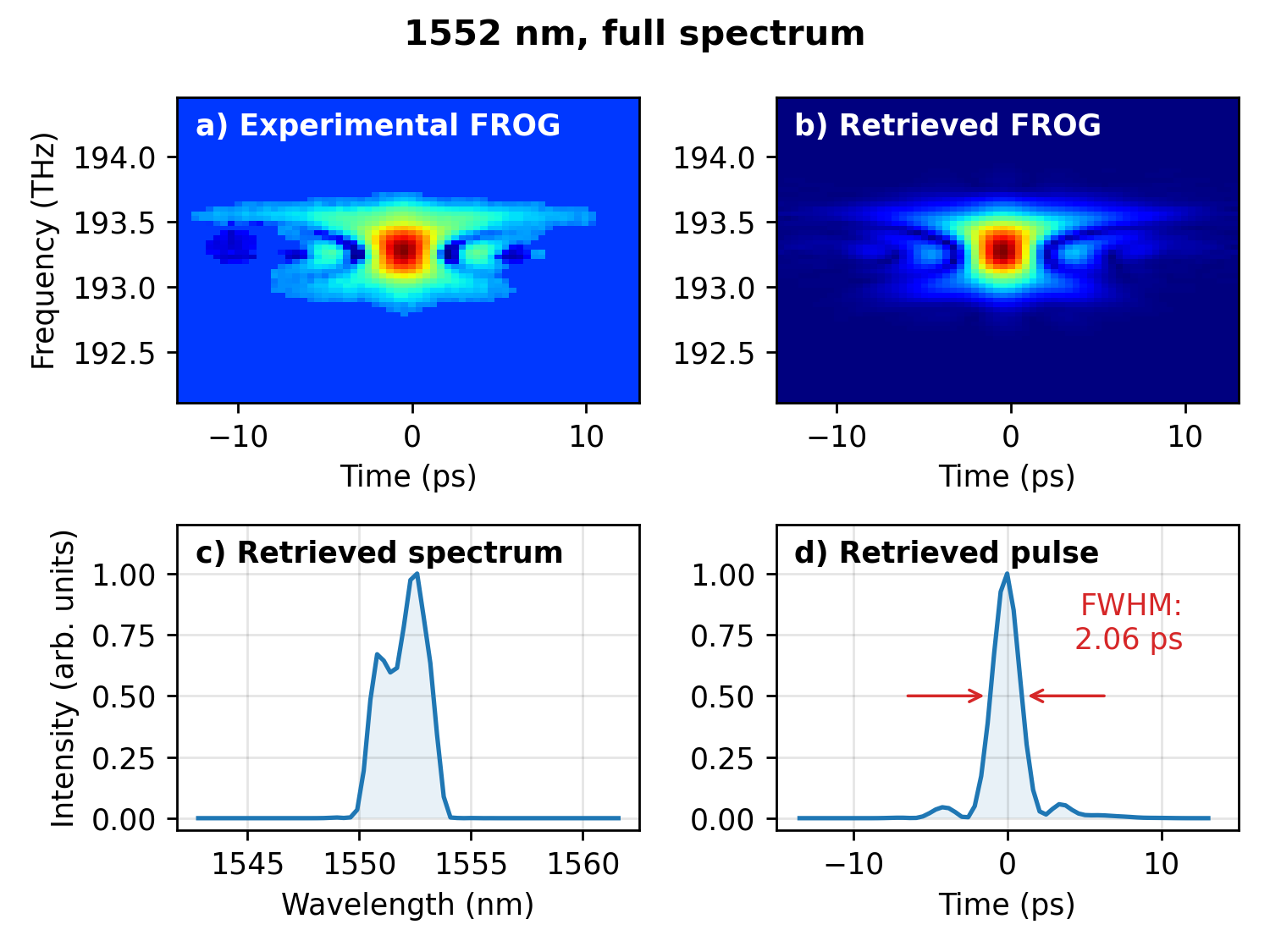}
\includegraphics[width=2.5in]{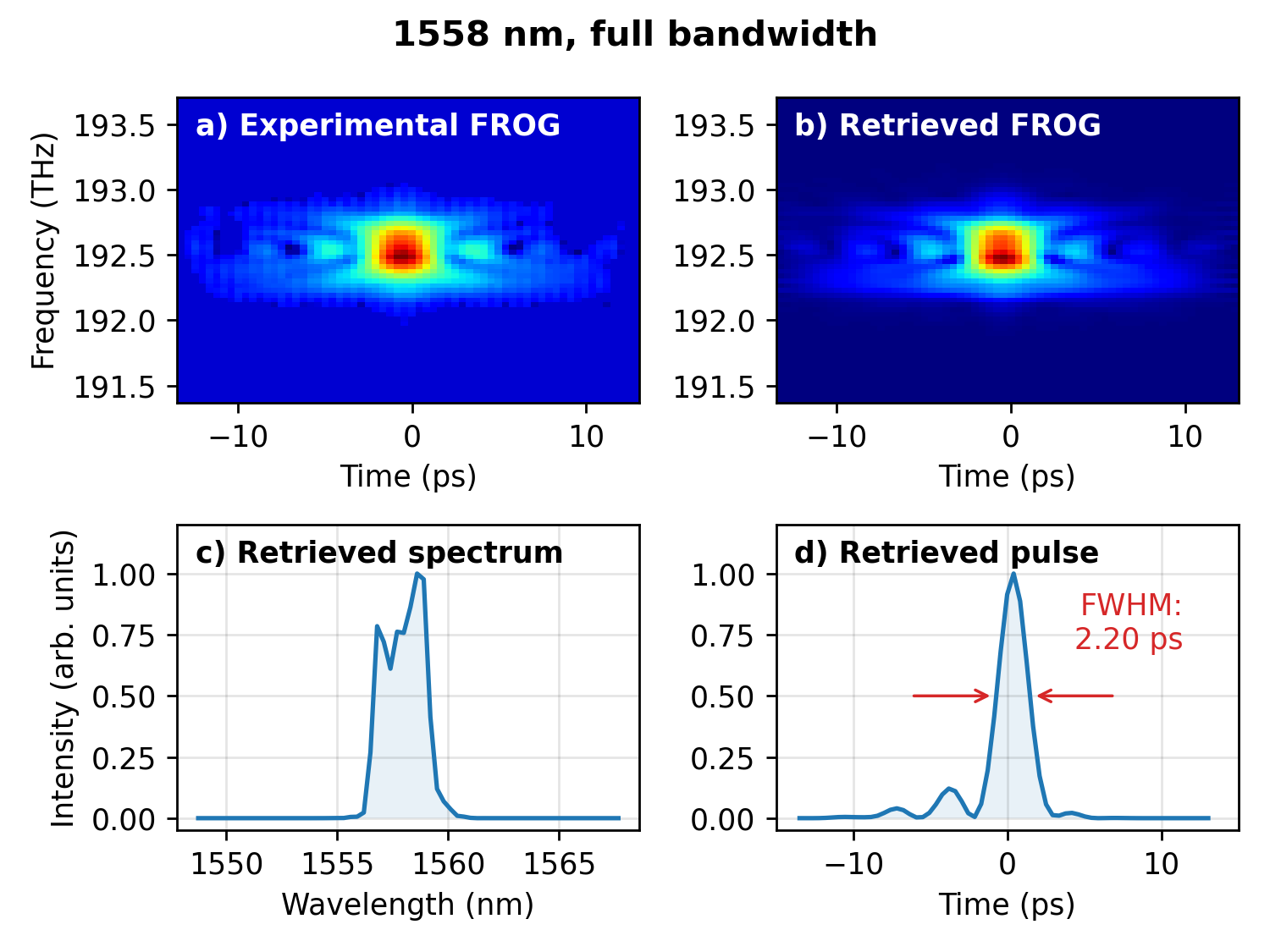}
\caption{\label{fig:frog_full} Pulse durations for the 1552 nm (left) and 1558~nm combs (right) measured with the frequency-resolved optical gating (FROG) technique. a) The experimentally recorded FROG spectrograms. b) The FROG spectrogram recovered using the phase-retrieval algorithm. c) The retrieved spectrum of the pulse as a function of wavelength. d) The retrieved pulse shape as a function of time. The pulse durations, as measured at the full-width-at-half-maximum (FWHM), are approximately 2~ps.}
\end{figure}

Pulse durations were measured using the frequency-resolved optical gating (FROG) technique \cite{trebino1997} using a MesaPhotonics VideoFROG system. The FROG spectrograms for 1558 and 1552 nm combs at their full bandwidth are shown in Figure~\ref{fig:frog_full}. In this case, the pulse durations are approximately 2~ps and the pulses exhibit a single peak in the time domain. 

\begin{figure}[ht!]
\centering\includegraphics[width=2.5in]{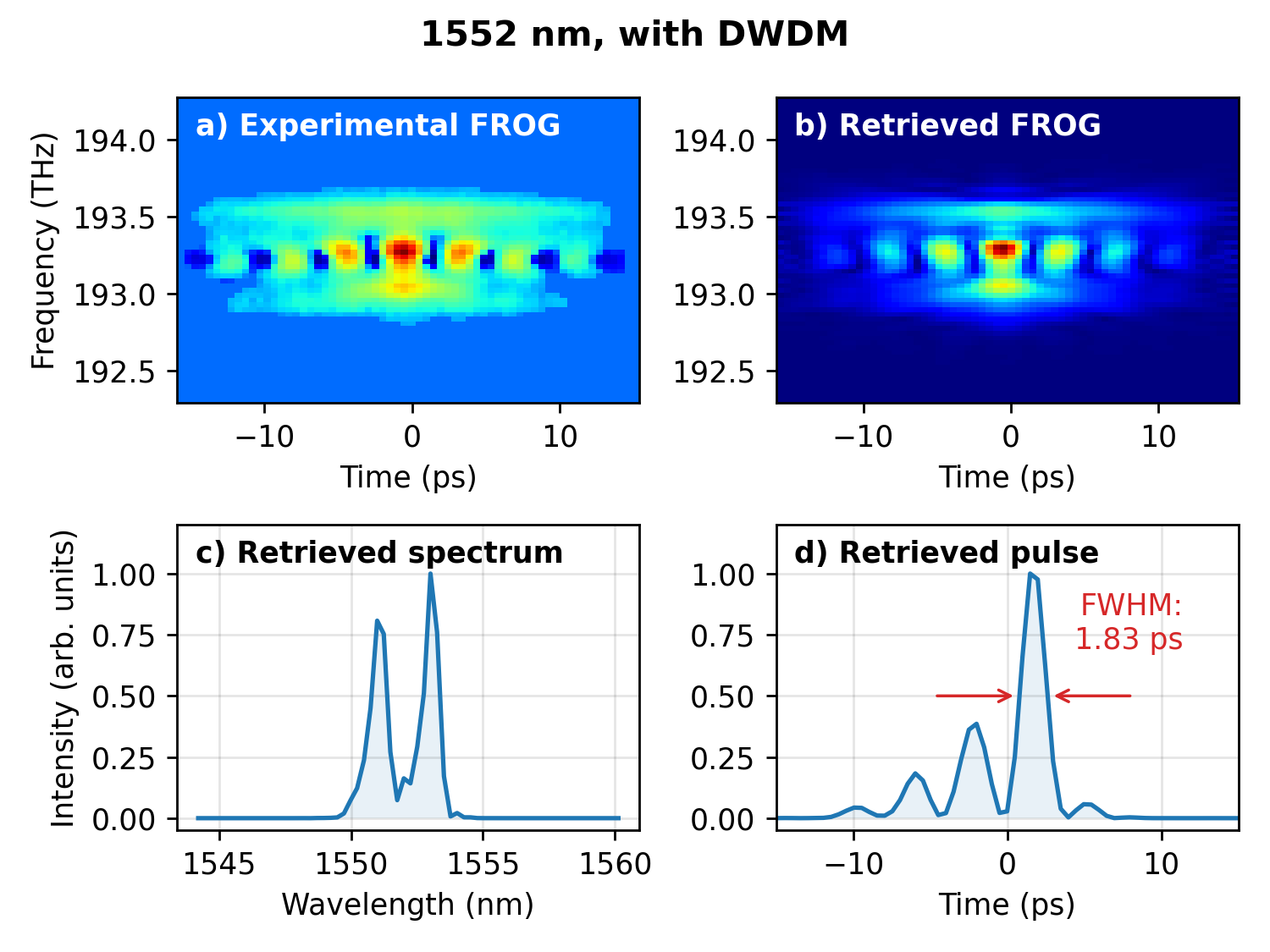}
\includegraphics[width=2.5in]{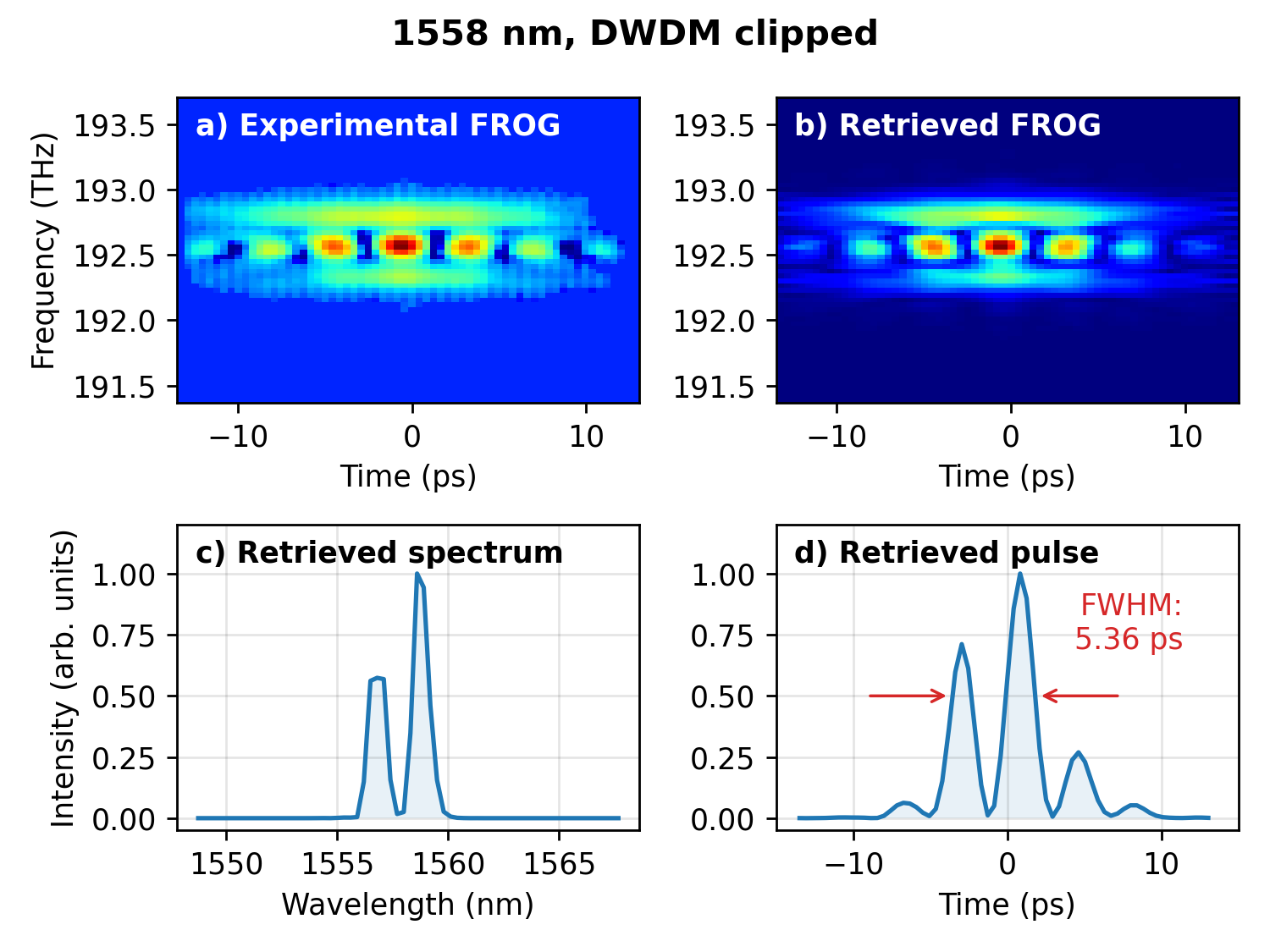}
\caption{\label{fig:frog_dwdm} Pulse durations for the 1552 nm (left) and 1558~nm combs (right) after the spectrum is modified by the DWDM. The pulses exhibit additional temporal structure.}
\end{figure}

When the DWDM is used to remove the center wavelengths, the pulse exhibits additional structure in the time domain. The resulting pulses still exhibit a center peak with a duration of approximately 2 ps, since the edge-to-edge bandwidth of the pulses is not changed. However, since the spectral character of the comb now consists of two peaks, the temporal structure of the pulses now consists of a pulse train of several pulses (Figure~\ref{fig:frog_dwdm}). For our purposes, it is sufficient to consider these structured pulses with a duration of roughly 5~ps. These pulses are still quite short compared to the round-trip time of the cavity (167~ps). 

A FROG trace was not recorded for the 1530~nm laser since our high-power laser amplifier did not provide sufficient gain in this wavelength range to achieve the required pulse energy to operate the FROG device. So, we cannot make claims about the pulse duration of the 1530~nm comb with certainty. However, since we used similar wavelengths to align the grating compressor, we anticipate that the pulses will still be close to their point of optimal compression. So, while the pulse durations may be somewhat longer with the 1530-nm comb, the pulses are most likely still short compared to the cavity round-trip time. 

\subsection{Shift versus power}
\begin{figure}[ht!]
\centering\includegraphics[width=4in]{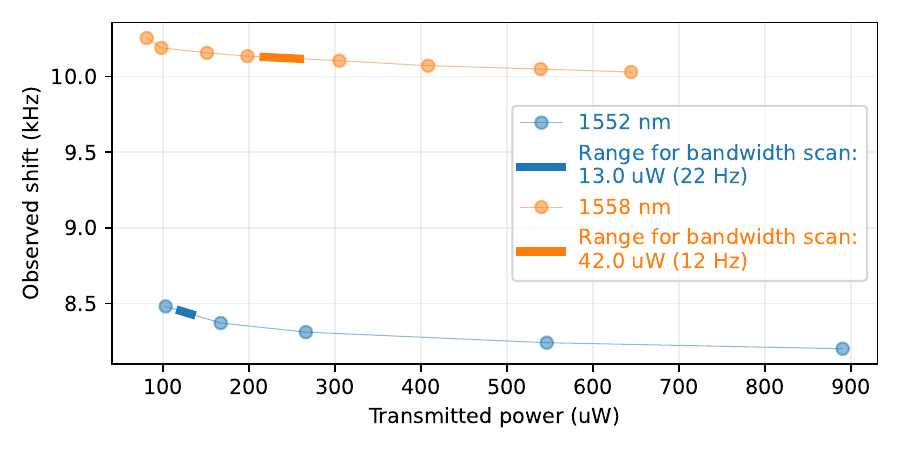}
\caption{\label{fig:power} Dependence of the experimentally observed shift on the power of the pulsed laser transmitted through the optical cavity. While there is a dependence of the shift on the power, the effect is small and mainly at lower powers.}
\end{figure}

It is possible that the average power of the CW and pulsed lasers could cause systematic errors with our measurements, potentially through the heating of the optical cavity. To determine the effect of varying the power, we measured how the pulsed-vs-cw shift changed at various input power levels, both at 1552 nm and 1558 nm  (Figure~\ref{fig:power}). Some small changes are seen, especially at lower powers, which we attribute to changes in the magnitude of the PDH signal, and the fact that the signal might be shifted slightly from zero volts.

Nevertheless, we do not expect that power changes had a significant effect on our measurements. The range used for the bandwidth scan presented in Figure~\ref{fig:shift_vs_bandwidth} is indicated in Figure~\ref{fig:power} and shows that, while there were small power changes during the bandwidth scan, these power changes will have a very small influence on the observed pulsed-vs-CW shift.

\vfil\eject

\bibliography{gravityCavity}

\begin{thebibliography}{10}
\newcommand{\enquote}[1]{``#1''}

\bibitem{einstein1907}
A.~Einstein, \enquote{On the inertia of energy required by the relativity principle,} {\protect\JournalTitle{Annalen der Physik}} \textbf{23}, 371--384 (1907).

\bibitem{einstein1912}
A.~Einstein, \enquote{{Gibt es eine Gravitationswirkung, die der Electrodynamischen Induktions\-wirkung Analog ist?}} {\protect\JournalTitle{Vier. Gericht. Med. und Sanit.}} \textbf{44}, 37--40 (1912).

\bibitem{einstein1915}
A.~Einstein and Grundgedanken, \enquote{der allgemeinen relativitätstheorie und anwendung dieser theorie in der astronomie,} {\protect\JournalTitle{Preuss. Akad. der Wissenschaften}} \textbf{315}, 778–786 (1915).

\bibitem{chou2010}
C.~W. Chou, D.~B. Hume, T.~Rosenband, and D.~J. Wineland, \enquote{Optical clocks and relativity,} {\protect\JournalTitle{Science}} \textbf{329}, 1630--1633 (2010).

\bibitem{herrmann2018}
S.~Herrmann, F.~Finke, M.~L\"ulf, O.~Kichakova, D.~Puetzfeld, D.~Knickmann, M.~List, B.~Rievers, G.~Giorgi, C.~G\"unther, H.~Dittus, R.~Prieto-Cerdeira, F.~Dilssner, F.~Gonzalez, E.~Sch\"onemann, J.~Ventura-Traveset, and C.~L\"ammerzahl, \enquote{Test of the gravitational redshift with galileo satellites in an eccentric orbit,} {\protect\JournalTitle{Phys. Rev. Lett.}} \textbf{121}, 231102 (2018).

\bibitem{bothwell2022}
T.~Bothwell, C.~Kennedy, A.~Aeppli, and et~al., \enquote{Resolving the gravitational redshift across a millimetre-scale atomic sample,} {\protect\JournalTitle{Nature}} \textbf{602}, 420–424 (2022).

\bibitem{shapiro1964}
I.~I. Shapiro, \enquote{{Fourth Test of General Relativity},} {\protect\JournalTitle{Phys. Rev. Lett.}} \textbf{13}, 789--791 (1964).

\bibitem{shapiro1968}
I.~I. Shapiro, \enquote{{Fourth Test of General Relativity: Preliminary Results},} {\protect\JournalTitle{Phys. Rev. Lett.}} \textbf{20}, 1265--1269 (1968).

\bibitem{Demorest_2010}
P.~B. Demorest, T.~Pennucci, S.~M. Ransom, M.~S.~E. Roberts, and J.~W.~T. Hessels, \enquote{{Shapiro Delay Measurement of a Two Solar Mass Neutron Star},} {\protect\JournalTitle{Nature}} \textbf{467}, 1081--1083 (2010). {Available at: \href{https://arxiv.org/abs/1010.5788} {arxiv.org/abs/1010.5788}}.

\bibitem{dyson1920}
F.~W. Dyson, A.~S. Eddington, and C.~Davidson, \enquote{{A Determination of the Deflection of Light by the Sun's Gravitational Field, from Observations Made at the Total Eclipse of May 29, 1919},} {\protect\JournalTitle{Philos. Trans. R. Soc. London, Ser. A}} \textbf{220 issue 571--581}, 291--333 (1920).

\bibitem{Einstein1922}
A.~Einstein, \emph{The Meaning of Relativity: Four Lectures Delivered at Princeton University, May, 1921} (Methuen \& Company Limited, 1922).

\bibitem{Thorne_1994}
K.~Thorne, \emph{Black Holes \& Time Warps: Einstein's Outrageous Legacy (Commonwealth Fund Book Program)} (WW Norton and Company, 1994).

\bibitem{mead2023}
C.~Mead, \enquote{Engineering view of gravitation,}  (2023). {Available at: \href{https://resolver.caltech.edu/CaltechAUTHORS:20230321-161151363} {resolver.caltech.edu/CaltechAUTHORS:20230321-161151363}}.

\bibitem{mead2023b}
C.~Mead, \enquote{A simple cosmology in g4v,} {\protect\JournalTitle{Symmetry}} \textbf{15} (2023).

\bibitem{Riess_2011}
{Riess, A. G., {\rm et al.}}, \enquote{{A 3\% Solution: Determination of the Hubble Constant with the Hubble Space Telescope and Wide Field Camera 3},} {\protect\JournalTitle{Astrophys. J.}} \textbf{730}, 119--174 (2011). {Available at: \href{https://arxiv.org/abs/1103.2976} {arxiv.org/abs/1103.2976}}.

\bibitem{kobayashi1972}
T.~Kobayashi, T.~Sueta, Y.~Cho, and Y.~Matsuo, \enquote{{High‐repetition‐rate optical pulse generator using a Fabry‐Perot electro‐optic modulator},} {\protect\JournalTitle{Applied Physics Letters}} \textbf{21}, 341--343 (1972).

\bibitem{kourogi1993}
M.~Kourogi, K.~Nakagawa, and M.~Ohtsu, \enquote{Wide-span optical frequency comb generator for accurate optical frequency difference measurement,} {\protect\JournalTitle{IEEE Journal of Quantum Electronics}} \textbf{29}, 2693--2701 (1993).

\bibitem{carlson2018}
D.~R. Carlson, D.~D. Hickstein, W.~Zhang, A.~J. Metcalf, F.~Quinlan, S.~A. Diddams, and S.~B. Papp, \enquote{Ultrafast electro-optic light with subcycle control,} {\protect\JournalTitle{Science}} \textbf{361}, 1358--1363 (2018).

\bibitem{treacy1969}
E.~Treacy, \enquote{Optical pulse compression with diffraction gratings,} {\protect\JournalTitle{IEEE Journal of Quantum Electronics}} \textbf{5}, 454--458 (1969).

\bibitem{trebino1997}
R.~Trebino, K.~W. DeLong, D.~N. Fittinghoff, J.~N. Sweetser, M.~A. Krumbügel, B.~A. Richman, and D.~J. Kane, \enquote{{Measuring ultrashort laser pulses in the time-frequency domain using frequency-resolved optical gating},} {\protect\JournalTitle{Review of Scientific Instruments}} \textbf{68}, 3277--3295 (1997).

\bibitem{siegman1986}
A.~E. Siegman, \emph{Lasers} (University Science Books, 1986).

\bibitem{drever1983}
R.~W.~P. Drever, J.~L. Hall, F.~V. Kowalski, J.~Hough, G.~M. Ford, A.~J. Munley, and H.~Ward, \enquote{Laser phase and frequency stabilization using an optical resonator,} {\protect\JournalTitle{Applied Physics B}} \textbf{2}, 97--105 (1983).

\bibitem{thorpe2005}
M.~J. Thorpe, R.~J. Jones, K.~D. Moll, J.~Ye, and R.~Lalezari, \enquote{Precise measurements of optical cavity dispersion and mirror coating properties via femtosecond combs,} {\protect\JournalTitle{Opt. Express}} \textbf{13}, 882--888 (2005).

\bibitem{paschotta2023}
R.~Paschotta, \enquote{Group delay dispersion,} {\protect\JournalTitle{RP Photonics Encyclopedia}}  (2023).

\bibitem{roslund2024}
J.~Roslund, A.~Cingöz, W.~Lunden, and et~al., \enquote{Optical clocks at sea,} {\protect\JournalTitle{Nature}} \textbf{628}, 736–740 (2024).

\bibitem{kedar2023}
D.~Kedar, J.~Yu, E.~Oelker, A.~Staron, W.~R. Milner, J.~M. Robinson, T.~Legero, F.~Riehle, U.~Sterr, and J.~Ye, \enquote{Frequency stability of cryogenic silicon cavities with semiconductor crystalline coatings,} {\protect\JournalTitle{Optica}} \textbf{10}, 464--470 (2023).

\bibitem{milner2019}
W.~R. Milner, J.~M. Robinson, C.~J. Kennedy, T.~Bothwell, D.~Kedar, D.~G. Matei, T.~Legero, U.~Sterr, F.~Riehle, H.~Leopardi, T.~M. Fortier, J.~A. Sherman, J.~Levine, J.~Yao, J.~Ye, and E.~Oelker, \enquote{Demonstration of a timescale based on a stable optical carrier,} {\protect\JournalTitle{Phys. Rev. Lett.}} \textbf{123}, 173201 (2019).

\bibitem{zhang2017}
W.~Zhang, J.~M. Robinson, L.~Sonderhouse, E.~Oelker, C.~Benko, J.~L. Hall, T.~Legero, D.~G. Matei, F.~Riehle, U.~Sterr, and J.~Ye, \enquote{Ultrastable silicon cavity in a continuously operating closed-cycle cryostat at 4 k,} {\protect\JournalTitle{Phys. Rev. Lett.}} \textbf{119}, 243601 (2017).

\end{thebibliography}

\end{document}